\documentclass[twocolumn,english,aps,pra,superscriptaddress]{revtex4-2}
\usepackage[T1]{fontenc}
\usepackage[utf8]{inputenc}
\usepackage{babel}
\usepackage{amsmath}
\usepackage{amssymb}
\usepackage{graphicx}
\newcommand{\rev}[1]{#1}
\usepackage[pdfusetitle,
 bookmarks=true,bookmarksnumbered=false,bookmarksopen=false,
 breaklinks=false,pdfborder={0 0 1},backref=false,colorlinks=false]
 {hyperref}
\begin{document}
\title{Thermal pseudo-transitions in a frustrated spin-pseudospin sawtooth
chain}
\author{Onofre Rojas}
\affiliation{Department of Physics, Institute of Natural Science, Federal University
of Lavras, Lavras, MG, Brazil}
\author{Jozef Strečka}
\affiliation{Department of Theoretical Physics and Astrophysics, Faculty of Science,
P. J. Šafárik University, Park Angelinum 9, 04001 Košice, Slovak Republic}
\begin{abstract}
We present an exact analysis of a spin--pseudospin X sawtooth chain
that incorporates three distinct valence states of copper ions and
serves as a minimal model for one-dimensional cuprates composed of
corner-sharing copper triangles. The model includes magnetic exchange
and electrostatic coupling constants both along the base sites and
between base and apex sites capturing the competition between spin
and charge degrees of freedom. Using the transfer-matrix method, we
derive exact expressions for the free energy and obtain an analytical
condition defining the pseudo-critical line associated with pronounced
thermodynamic anomalies. The ground-state analysis reveals, besides
an antiferromagnetic phase, three frustrated phases characterized
by distinct residual entropies. At finite temperatures, these zero-temperature
phase boundaries evolve into narrow entropy ridges signaling pseudo-transitions
between the corresponding quasi-phases. 
The specific heat and the \rev{avoided-crossing scale exhibit sharp but 
continuous peaks at the pseudo-critical temperature, whereas the 
physical correlation length may be controlled by a different 
subleading eigenvalue}. Local correlation functions uncover a cooperative rearrangement
from charge-dominated to magnetically correlated regimes. Our results
demonstrate that the sawtooth geometry promotes frustration and short-range
coherence leading to pronounced pseudo-transition behavior. 
\end{abstract}
\maketitle

\section{Introduction}

One-dimensional lattice-statistical models have long served as a paradigmatic
framework for understanding cooperative phenomena in strongly correlated
electron and spin systems due to their exact solvability and conceptual
transparency~\citep{mattis}. Despite the absence of true finite-temperature
phase transitions in one-dimensional systems with short-range interactions
as established by rigorous non-existence theorems~\citep{vanhove,cuesta},
a growing class of exactly solvable models has been shown to exhibit
pseudo-transition phenomena. The pseudo-transitions refer to sharp
but continuous anomalies in thermodynamic quantities that closely
mimic genuine critical behavior. Such phenomena have been reported
in a variety of one-dimensional spin and electron systems, where competing
interactions give rise to a quasi-critical behavior and pronounced
short-range ordering effects \citep{galisova2015,rojaspseudo,universal,rojase,hutak2021,yin2024,yin2024prb}.

Among others, the pseudo-transition phenomena have also been observed
in geometrically frustrated sawtooth (delta) spin chains, where competing
interactions and frustrated lattice topology enhance a quasi-critical
behavior \citep{yin-tsvelik-prl-2024,yin-prb-2026}. In addition,
analogous pseudo-critical features have been reported in one-dimensional
strongly correlated electron systems demonstrating that such phenomena
extend beyond traditional spin models \citep{rojas-pre-2021,rojas-pre-2024}.
A particularly appealing platform for studying pseudo-transitions
is provided by spin--pseudospin models of cuprate chains, where magnetic
and charge degrees of freedom are intrinsically coupled. In the previous
work by one of the present authors~\citep{strecka_epjb}, a minimal
spin--pseudospin model was introduced for one-dimensional cuprate
chains $[\mathrm{CuO}]_{\infty}$ that demonstrated the emergence
of a pseudo-transition between charge-ordered and antiferromagnetic
regimes accompanied by sharp anomalies in entropy, specific heat,
and correlation functions. This approach captures the essential physics
of competing valence configurations of Cu$^{1+}$, Cu$^{2+}$, and
Cu$^{3+}$ ions and provides a simple yet powerful framework for analyzing
spin--charge interplay in low-dimensional systems~\citep{moskvin1,moskvin2,panov}.

In the present work, we extend this framework to a geometrically frustrated
spin--pseudospin model on a sawtooth (delta) chain composed of corner-sharing
triangular units. The introduction of geometric frustration fundamentally
alters the balance between magnetic exchange and electrostatic interactions
leading to a significantly richer ground-state structure compared
to the linear chain. In particular, the triangular geometry allows
for multiple competing configurations of spins and pseudospins, which
results in several distinct frustrated phases in addition to antiferromagnetically
ordered state. This enhanced degeneracy and competition provide a
natural setting for the emergence of more complex pseudo-transition
behavior and diverse quasi-phases in line with general expectations
for frustrated low-dimensional systems.

From an experimental perspective, frustrated sawtooth (delta) chains
are realized in several classes of low-dimensional cuprates. In particular,
some delafossite compounds MCuO$_{2+\delta}$ provide experimental
realizations of frustrated chains of corner-sharing triangles constituted
by copper ions, where competing exchange interactions and charge fluctuations
play a crucial role~\citep{delafossite1,delafossite2,delafossite3,delafossite4,delafossite5}.
These materials exhibit a rich interplay between magnetism, charge
ordering, and lattice geometry making them promising candidates for
observing pseudo-transition phenomena and short-range correlated regimes
analogous to those predicted by the present model.

The aim of this paper is to provide an exact analytical treatment
of the spin--pseudospin sawtooth chain and to elucidate the role
of geometric frustration in shaping its thermodynamic and correlation
properties. By employing the transfer-matrix method, we derive closed-form
expressions for the partition function and relevant observables and
we identify the pseudo-critical conditions governing the crossover
between distinct quasi-phases. Our results demonstrate that frustration
not only enriches the ground-state phase diagram but also enhances
the signatures of pseudo-transitions, leading to sharper anomalies
and a more intricate interplay between spin and charge degrees of
freedom.

The paper is organized as follows. In Sec.~\ref{sec:Model} we introduce
the one-dimensional spin-pseudospin model that describes sawtooth
chain composed of corner-sharing triangles and present the transfer-matrix
formulation and pseudo-critical condition. Sec.~\ref{sec:Thermod}
reports comprehensive analysis of the ground-state phase diagram and
quasi-phases. Sec.~\ref{sec:Results} discusses most intriguing results
for thermodynamic quantities including entropy, specific heat, correlation
length, and the local correlations. Finally, we summarize our main
findings and physical implications of the study in Sec.~\ref{sec:Conclusion}.

\section{Model and method}

\label{sec:Model}

\begin{figure}
\includegraphics[scale=0.33]{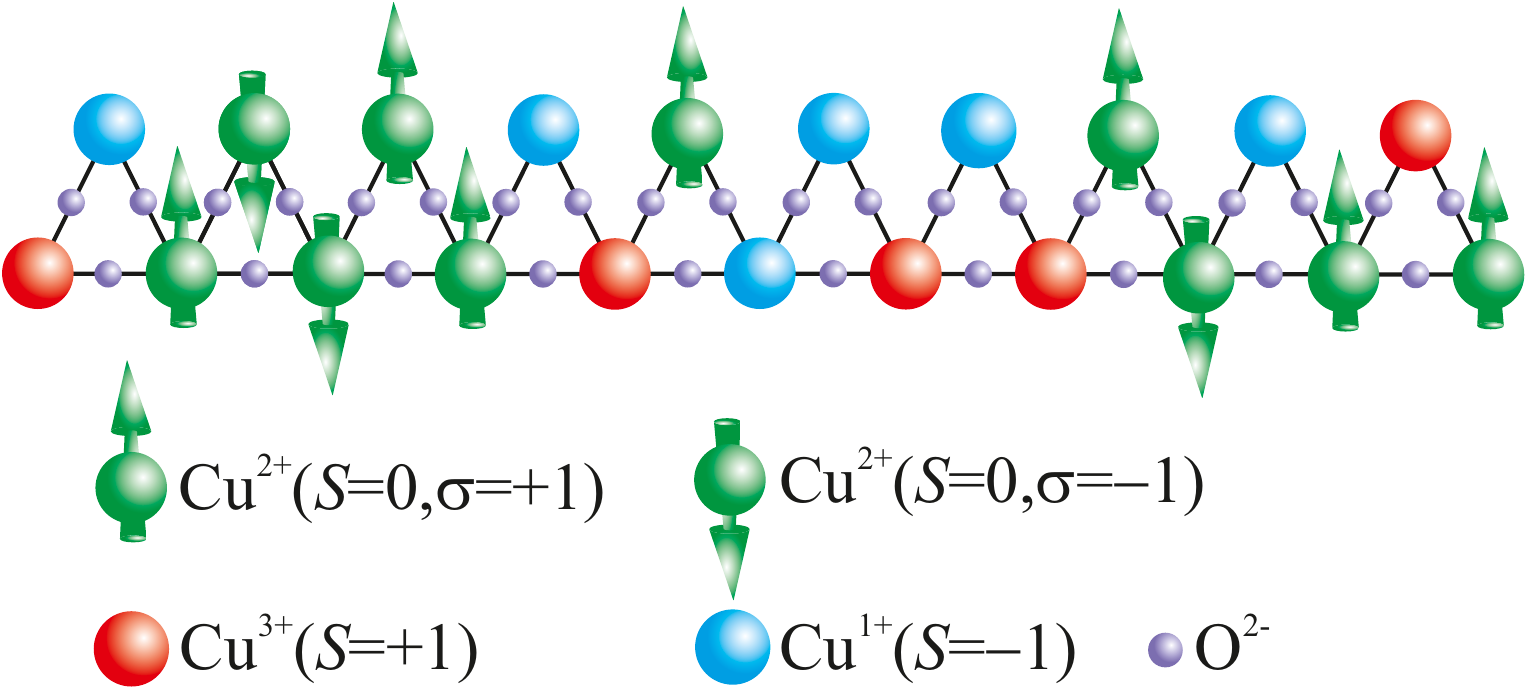}\caption{A schematic illustration of the spin-pseudospin sawtooth chain and
one of its allowed configurations.}
\label{fig:saw} 
\end{figure}

We consider one-dimensional sawtooth chain composed of alternating
base and apex sites that form a sequence of corner-sharing triangular
units as schematically illustrated in Fig. \ref{fig:saw}. Each triangular
cell constitutes the minimal structural motif capable of capturing
the competition between magnetic and charge degrees of freedom characteristic
of cuprate chains emergent in the delafossite family \citep{delafossite1,delafossite2,delafossite3,delafossite4,delafossite5}.
To describe these intertwined degrees of freedom, we associate with
each lattice site a pair of variables - an Ising spin and a pseudospin
- representing the magnetic and charge degrees of freedom, respectively.
The base sites are characterized by spin-pseudospin pairs $\sigma_{i}$
and $S_{i}$, whereas the apex sites are described by the spin-pseudospin
pairs $\tilde{\sigma}_{i}$ and $\tilde{S}_{i}$. The pseudospin variables
take values $S_{i},\tilde{S}_{i}\in\{-1,0,1\}$ corresponding to the
three nominal valence states of Cu$^{1+}$, Cu$^{2+}$, Cu$^{3+}$
centers. The charged pseudospin values $S_{i},\tilde{S}_{i}=-1$ and
$+1$ represent the negatively and positively charged valence states
Cu$^{1+}$ and Cu$^{3+}$ when considering the overall electrostatic
balance with respect to O$^{2-}$ anion, while the pseudospin value
$S_{i},\tilde{S}_{i}=0$ corresponds to the neutral valence state
Cu$^{2+}$ described by the Ising spin $\sigma_{i},\tilde{\sigma}_{i}\in\{\pm1\}$
representing two possible magnetic states of the spin-1/2 Cu$^{2+}$
centers.

\subsection{Spin-pseudospin sawtooth chain: Hamiltonian}

The total Hamiltonian of the spin-pseudospin sawtooth chain can be
for further convenience expressed as a sum over local Hamiltonians
of individual triangular unit cells 
\begin{equation}
{\cal H}=\sum^{N}_{i=1}{\cal H}_{i,i+1},
\end{equation}
where each local cell Hamiltonian is given by 
\begin{alignat}{1}
{\cal H}_{i,i+1}= & J_{1}(1-\tilde{S}^{2}_{i})\tilde{\sigma}_{i}\left[(1-S^{2}_{i})\sigma_{i}+(1-S^{2}_{i+1})\sigma_{i+1}\right]\nonumber \\
 & +J(1-S^{2}_{i})\sigma_{i}(1-S^{2}_{i+1})\sigma_{i+1}\nonumber \\
 & +V_{1}\tilde{S}_{i}\left(S_{i}+S_{i+1}\right)+VS_{i}S_{i+1}\nonumber \\
 & +\frac{\Delta}{2}(S^{2}_{i}+S^{2}_{i+1})+\Delta_{1}\tilde{S}^{2}_{i}.\label{eq:H-cell}
\end{alignat}
The coupling constants $J>0$ and $J_{1}>0$ determine the antiferromagnetic
exchange interactions between spins on the base--base and base--apex
bonds, respectively. Analogously, the electrostatic potentials $V$
and $V_{1}$ quantify the energy cost associated with presence of
the charged particles on the respective bonds. Finally, the on-site
terms $\Delta$ and $\Delta_{1}$ act as chemical potentials controlling
the energy cost associated with creating charged states on the base
and apex sites, respectively, and thus regulate the equilibrium densities
of charged carriers. The sawtooth chain consists of $N$ identical
triangular units with periodic boundary conditions assumed.

\subsection{Transfer-matrix solution}\label{subsec:TMs}

The statistical properties of the spin-pseudospin sawtooth chain can
be exactly evaluated by means of the transfer-matrix method \citep{baxt82}.
Within this formalism, the partition function can be expressed as
\begin{equation}
Z=\sum_{\{S,\sigma\}}\prod^{N}_{i=1}\mathbf{T}(S_{i},\sigma_{i};S_{i+1},\sigma_{i+1}),
\end{equation}
where the transfer matrix $\mathbf{T}(S_{i},\sigma_{i};S_{i+1},\sigma_{i+1})$
explicitly depends on two consecutive spin-pseudospin pairs from adjacent
base sites. This transfer matrix is obtained from the Bolztmann weight
associated with the cell Hamiltonian (\ref{eq:H-cell}) by tracing
out degrees of freedom of the apex spin $\tilde{\sigma}_{i}$ and
pseudospin $\tilde{S}_{i}$ 
\begin{equation}
\mathbf{T}(S_{i},\sigma_{i};S_{i+1},\sigma_{i+1})=\sum^{1}_{\tilde{S}_{i}=-1}\sum_{\tilde{\sigma}_{i}=\pm1}\frac{{\rm e}^{-\beta\mathcal{H}_{i,i+1}}}{\sqrt{2^{S^{2}_{i}+S^{2}_{i+1}}}}.\label{tm}
\end{equation}
The trace over the degrees of freedom of this spin-pseudospin pair
can be carried out analytically yielding 
\begin{alignat*}{1}
\mathbf{T}(S_{i},\sigma_{i};S_{i+1},\sigma_{i+1})= & \frac{2\mathrm{e}^{-\beta E_{{\rm b}}}}{\sqrt{2^{S^{2}_{i}+S^{2}_{i+1}}}}\Big[\cosh\big(\beta J_{1}\tau_{i,i+1}\big)+\\
 & +\rev{2\mathrm{e}^{-\beta\Delta_{1}}}\cosh\big(\beta V_{1}(S_{i}+S_{i+1})\big)\Big],
\end{alignat*}
where 
\begin{alignat*}{1}
E_{{\rm b}}\!=\!J(1\!-\!S^{2}_{i})(1\!-\!S^{2}_{i+1})\sigma_{i}\sigma_{i+1}\!+\!\tfrac{\Delta}{2}(S^{2}_{i}\!+\!S^{2}_{i+1})\!+\!VS_{i}S_{i+1},
\end{alignat*}
and $\tau_{i,i+1}=(1-S^{2}_{i})\sigma_{i}+(1-S^{2}_{i+1})\sigma_{i+1}$.
A spin-pseudospin pair ascribed to the same base
site acquires six possible combinations, which can be ordered in the
transfer matrix as $(S,\sigma)=\{(1,1),(1,-1),(0,1),(0,-1),(-1,1),(-1,-1)\}$
\begin{equation}
\mathbf{T}=\left(\begin{array}{cccccc}
w_{1} & w_{1} & w_{2} & w_{2} & w_{5} & w_{5}\\
w_{1} & w_{1} & w_{2} & w_{2} & w_{5} & w_{5}\\
w_{2} & w_{2} & w_{3} & w_{4} & w_{2} & w_{2}\\
w_{2} & w_{2} & w_{4} & w_{3} & w_{2} & w_{2}\\
w_{5} & w_{5} & w_{2} & w_{2} & w_{1} & w_{1}\\
w_{5} & w_{5} & w_{2} & w_{2} & w_{1} & w_{1}
\end{array}\right).\label{tmee}
\end{equation}
The five independent Boltzmann weights $w_{1},\dots,w_{5}$ follow
directly from the general expression defining the transfer matrix
(\ref{tm}), which depends on the configurations of the spin-pseudospin
pairs associated with the base sites and can be written explicitly
as 
\begin{alignat}{1}
w_{1}= & \mathrm{e}^{-\beta(\Delta+V)}\Big[1+2\mathrm{e}^{-\beta\Delta_{1}}\cosh(2\beta V_{1})\Big],\nonumber \\
w_{2}= & \sqrt{2}\mathrm{e}^{-\beta\Delta/2}\Big[\cosh(\beta J_{1})+2\mathrm{e}^{-\beta\Delta_{1}}\cosh(\beta V_{1})\Big],\nonumber \\
w_{3}= & 2\mathrm{e}^{-\beta J}\Big[\cosh(2\beta J_{1})+2\mathrm{e}^{-\beta\Delta_{1}}\Big],\nonumber \\
w_{4}= & 2\mathrm{e}^{\beta J}\Big[1+2\mathrm{e}^{-\beta\Delta_{1}}\Big],\nonumber \\
w_{5}= & \mathrm{e}^{-\beta(\Delta-V)}\Big[1+2\mathrm{e}^{-\beta\Delta_{1}}\Big].\label{tme}
\end{alignat}
In the limiting case $J_{1}=\Delta_{1}=V_{1}=0$, the Boltzmann weights
(\ref{tme}) reduce to those of the one-dimensional spin-pseudospin
linear chain studied previously in Ref. \citep{strecka_epjb}.

It follows from Eq. (\ref{tmee}) that the rows and columns of the
transfer matrix occur in three identical pairs, namely (1,2), (3,4),
and (5,6), which enables its reduction into symmetric subspaces. Exploiting
this block structure, each pair can be decomposed into symmetric and
antisymmetric linear combinations. The transfer matrix has two zero
eigenvalues obtained upon eliminating the antisymmetric basis vectors
associated with the pairs $(1,2)$ and $(5,6)$. The remaining four
nonzero eigenvalues belong to two distinct subspaces:

(i) The three-dimensional symmetric subspace spanned by the three
pairs. Within this subspace, a convenient orthonormal basis is constructed
from the pair-averaged combinations of $(1,2)$, $(3,4)$, $(5,6)$.
In this basis, the transfer matrix reduces to the $3\times3$ block
\begin{equation}
\mathbf{T}_{s}=\begin{pmatrix}2w_{1} & 2w_{2} & 2w_{5}\\
2w_{2} & w_{3}+w_{4} & 2w_{2}\\
2w_{5} & 2w_{2} & 2w_{1}
\end{pmatrix}.\label{eq:TmR}
\end{equation}
This $3\times3$ matrix has one obvious eigenvector proportional to
$(1,0,-1)$ yielding the eigenvalue $\lambda_{3}=2\,(w_{1}-w_{5})$.
The remaining two eigenvalues are obtained from the $2\times2$ sub-block,
which is defined by restricting $\mathbf{T}_{s}$ within the subspace
spanned by the vectors $(1,0,1)$ and $(0,1,0)$. 
\begin{alignat}{1}
\lambda_{1}= & \frac{1}{2}\left[v_{1}+v_{2}+\sqrt{\left(v_{1}-v_{2}\right)^{2}+32w^{2}_{2}}\right],\label{eq:l1}\\
\lambda_{2}= & \frac{1}{2}\left[v_{1}+v_{2}-\sqrt{\left(v_{1}-v_{2}\right)^{2}+32w^{2}_{2}}\right],\label{eq:l2}\\
\lambda_{3}= & 4\mathrm{e}^{-\beta\Delta}\Big[2\mathrm{e}^{-\beta(\Delta_{1}+V)}\sinh^{2}(\beta V_{1})\nonumber \\
 & -\big(1+2\mathrm{e}^{-\beta\Delta_{1}}\big)\sinh(\beta V)\Bigr],\label{eq:l3}
\end{alignat}
where $v_{1}=2(w_{1}+w_{5})$ and $v_{2}=w_{3}+w_{4}$.

(ii) The remaining eigenstate lies in the antisymmetric subspace associated
with the pair (3,4). In this subspace the antisymmetric combination
is an eigenvector of the transfer matrix with the eigenvalue 
\begin{alignat}{1}
\lambda_{4}= & w_{3}-w_{4}\\
\lambda_{4}= & 4\mathrm{e}^{-\beta J}\sinh^{2}(\beta J_{1})-4\big(1+2\mathrm{e}^{-\beta\Delta_{1}}\big)\sinh(\beta J),
\end{alignat}

It can be easily verified from these results that the first eigenvalue
$\lambda_{1}$ is the dominant eigenvalue governing the thermodynamics
of the model, which uniquely determines the free energy per unit cell
in the thermodynamic limit $f=-\frac{1}{\beta}\ln(\lambda_{1})$.
Consequently, this eigenvalue incorporates the dominant statistical
weight of the spin-pseudospin sawtooth chain and fully determines
all thermodynamic observables discussed in the following sections.
Its explicit dependence on the coupling constants ($J$, $J_{1}$,
$V$, $V_{1}$, $\Delta$, $\Delta_{1}$) reflects the intricate interplay
between magnetic exchange interactions and charge correlations characteristic
of cuprate chains consisting of corner-sharing triangles.
\rev{Although $\lambda_2$ is the second-largest eigenvalue, it is not necessarily
the second largest in magnitude. Hence, the correlation length can be expressed
as $\xi=1/\ln(\lambda_1/\lambda_{\rm sub})$, where 
$\lambda_{\rm sub}=\max(|\lambda_2|,|\lambda_3|,|\lambda_4|)$.
The avoided-crossing quantity is instead given by
$\xi_{\rm ac}=1/\ln(\lambda_1/\lambda_2)$.}

\subsection{Pseudo-critical condition}

\label{subsec:psd-cirt-cond}

In analogy with other one-dimensional models displaying pseudo-transitions,
the crossover between distinct quasi-phases in the spin--pseudospin
sawtooth chain is governed by a simple analytical condition among
the Boltzmann weights. For the present geometry, this condition reads
\begin{equation}
w_{1}+w_{5}=\frac{w_{3}+w_{4}}{2},\label{eq:psd-cond}
\end{equation}
and defines the pseudo-critical line at which the most pronounced
anomalies in thermodynamic quantities, such as sharp peaks in the
specific heat and the avoided-crossing, occur. Expressed explicitly
in terms of the model parameters, the pseudo-critical condition takes
the form 
\begin{equation}
\mathrm{e}^{-\beta\Delta_{1}}=\frac{e^{\beta\Delta}(e^{-\beta J}\cosh(2\beta J_{1})+e^{\beta J})-2\cosh(\beta V)}{2\big(e^{-\beta V}\cosh(2\beta V_{1})+e^{\beta V}\big)-4e^{\beta\Delta}\cosh(\beta J)}.
\end{equation}
This relation determines the pseudo-critical temperature $T_{p}$
separating regimes dominated by charge and spin short-range correlations,
which provides an analogue of the boundary between charge-ordered
and antiferromagnetic regimes in $[\mathrm{CuO}]_{\infty}$ cuprate
chains~\citep{strecka_epjb}.

In the low-temperature limit, assuming all parameters positive and
setting $V_{1}=V$ and $\Delta_{1}=\Delta$, the pseudo-transition
condition reduces to a balance between the dominant exponential contributions
in the numerator and denominator. Retaining only these leading-order
terms, the relation can be written as 
\begin{equation}
6\mathrm{e}^{\beta(V-2\Delta)}+2\mathrm{e}^{\beta(V-\Delta)}=\mathrm{e}^{\beta(2J_{1}-J)}+4\mathrm{e}^{\beta(J-\Delta)}+2\mathrm{e}^{\beta J}.\label{eq:prt-Tp-cond}
\end{equation}
The formula~(\ref{eq:prt-Tp-cond}) is more amenable to analytical
treatment and provides a direct means of estimating the pseudo-critical
temperature $T_{p}$ as a function of the interaction parameters.
The resulting balance between the leading exponential terms determines
the approximate location of the pseudo-critical lines separating charge
and spin ordered quasi-phases.

\subsection{Spectral interpretation of the pseudo-transition}

The pseudo-critical condition derived in Sec. \ref{subsec:psd-cirt-cond}
admits a direct interpretation in terms of the spectral structure
of the transfer matrix introduced in Sec. \ref{subsec:TMs}. The thermodynamic
behavior is governed by the largest eigenvalue $\lambda_{1}$, while the 
eigenvalue $\lambda_2$ controls \rev{the avoided crossing between the 
two competing sectors, whereas the physical correlation length 
is determined by the subleading eigenvalue of largest magnitude.}
 From Eqs. (\ref{eq:l1}-\ref{eq:l2}),
the two largest eigenvalues differ only by the sign of the square-root
term, whose argument is $(v_{1}-v_{2})^{2}+32w^{2}_{2}$. This quantity
is minimized when $v_{1}=v_{2}$, which is equivalent to the pseudo-critical
condition (\ref{eq:psd-cond}).

According to \citep{spc-crit}, the dimensionless spectral splitting
can be written as
\begin{equation}
\zeta\sim\frac{\sqrt{(v_{1}-v_{2})^{2}+32w^{2}_{2}}}{v_{1}+v_{2}},
\end{equation}
showing explicitly that the relevant scale is set by the competition
between the contributions $v_{1}$ and $v_{2}$, while the parameter
$w_{2}$ provides a finite coupling that prevents exact degeneracy.

This structure follows directly from the reduced transfer matrix (\ref{eq:TmR}).
In this representation, the diagonal terms proportional to $v_{1}$
and $v_{2}$ define two dominant sectors, whereas the off-diagonal
term proportional to $w_{2}$ couples them. As a result, the transfer
matrix remains irreducible, but approaches a nearly block-diagonal
form in the regime $v_{1}\approx v_{2}$.

In this regime, the two sectors contribute comparably to the partition
function, and the leading eigenvalues become quasi-degenerate. This
realizes explicitly the general mechanism of pseudotransitions in
one-dimensional systems, where sharp but analytic anomalies arise
from the competition between weakly coupled sectors within an irreducible
transfer matrix \citep{spc-crit}.

\section{Ground states and quasi-phases}

\label{sec:Thermod}

In this section, we explore the thermodynamic behavior of the spin-pseudospin
sawtooth chain as obtained from the largest eigenvalue $\lambda_{1}$
of the transfer matrix. This eigenvalue fully determines the free
energy per unit cell and consequently, all relevant physical quantities
such as entropy, specific heat, and correlation functions. The sawtooth-chain
geometry with its alternating base and apex sites introduces an intrinsic
competition between magnetic and charge degrees of freedom that mimics
the local environment of cuprate chains composed from corner-sharing
triangles. By examining the temperature dependence of these thermodynamic
quantities, we identify the signatures of pseudo-transitions arising
from frustration and short-range ordering.

The following subsections present results for the entropy landscape,
specific heat anomalies, and correlation functions, providing a comprehensive
picture of how the interplay between exchange and electrostatic interactions
governs the emergence of quasi-phases in this geometrically frustrated
system.

\begin{figure}
\includegraphics[scale=0.28]{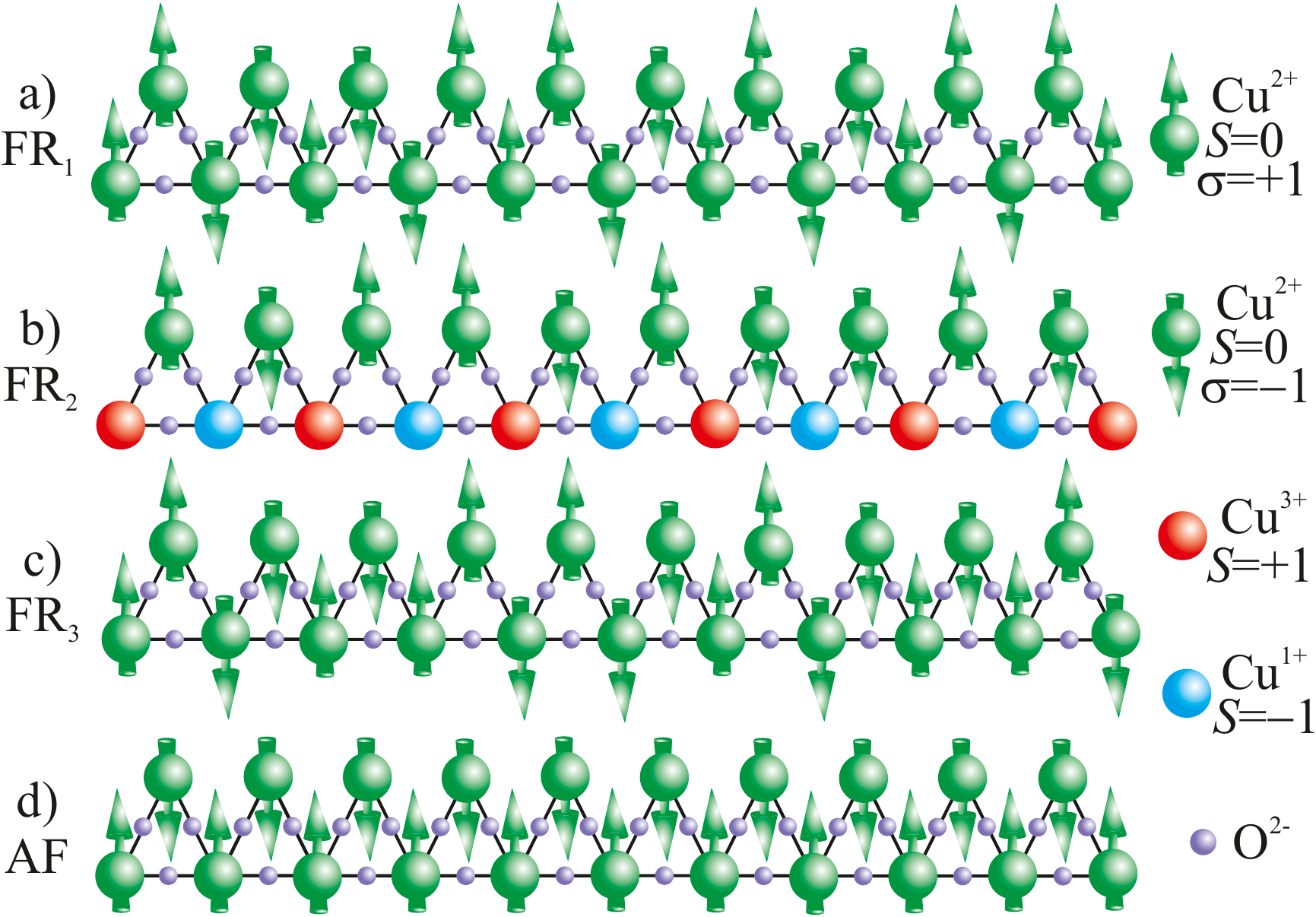}\caption{Four available ground states of the spin-pseudospin sawtooth chain
under the condition $\Delta_{1}=\Delta$ and $V_{1}=V$: (a) FR$_{1}$
phase with the Néel order of the base spins and a complete randomness
of the apex spins; (b) FR$_{2}$ phase with the charge order of the
base entities and a complete randomness of the apex spins; (c) FR$_{3}$
phase with a random three-fold degenerate character of each triangular
unit; (d) AF phase with the antiferromagnetic long-range order of
the base and apex spins.}
\label{fig:gs} 
\end{figure}

\subsection{Ground-state phase diagram}\label{subsec:sec.3.A}

\begin{figure*}
\centering \includegraphics{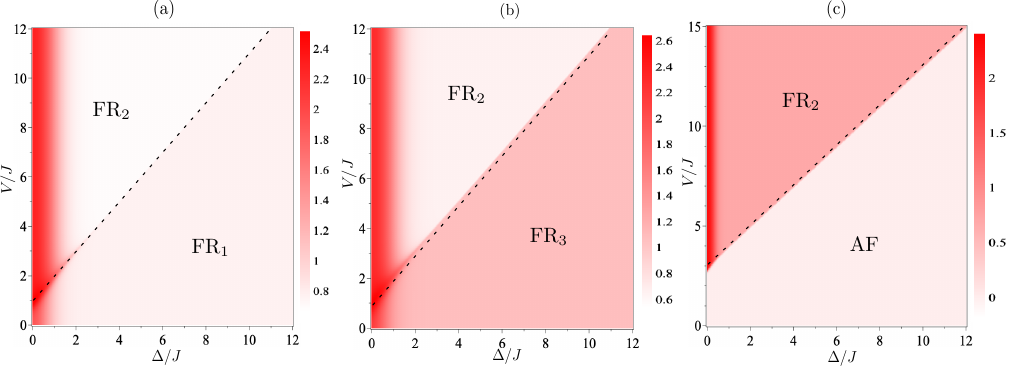}\caption{Ground-state phase diagram in the plane $V/J-\Delta/J$ assuming $\Delta_{1}=\Delta$
and $V_{1}=V$ for three distinct values of the coupling ratio: (a)
$J_{1}/J=0.5$, (b) $J_{1}/J=1$, (c) $J_{1}/J=2$. Background density
plot displays the entropy \rev{$\mathcal{S}/(Nk_{\mathrm{B}})$} at a fixed temperature
$k_{B}T/J=0.1$.}
\label{fig:Phase-diagram} 
\end{figure*}

Fig. \ref{fig:Phase-diagram} displays the ground-state phase diagram
of the spin-pseudospin sawtooth chain in the $V/J-\Delta/J$ plane
highlighting three frustrated phases and one antiferromagnetic phase.
Each phase can be characterized as a product state constructed from
the lowest-energy state of an individual unit cell. At low values
of $V/J$, the ground state corresponds to the first frustrated phase
${\rm FR}_{1}$, which incorporates Néel long-range order of the base
spins causing fully frustrated (paramagnetic) character of the apex
spins as illustrated in Fig. \ref{fig:gs}(a) and given by 
\begin{equation}
|{\rm FR}_{1}\rangle=\begin{cases}
\begin{alignedat}{1}\bigotimes^{N/2}_{i=1}\bigl|{}^{\tilde{S}_{2i-1}=0,\tilde{\sigma}_{2i-1}}_{S_{2i-1}=0,\sigma_{2i-1}=1}{}^{;\tilde{S}_{2i}=0,\tilde{\sigma}_{2i}}_{;S_{2i}=0,\sigma_{2i}=-1}\bigr\rangle,\\
\bigotimes^{N/2}_{i=1}\bigl|{}^{\tilde{S}_{2i-1}=0,\tilde{\sigma}_{2i-1}}_{S_{2i-1}=0,\sigma_{2i-1}=-1}{}^{;\tilde{S}_{2i}=0,\tilde{\sigma}_{2i}}_{;S_{2i}=0,\sigma_{2i}=1}\bigr\rangle.
\end{alignedat}
\end{cases}
\end{equation}
\rev{Here, the variables associated with a base site are displayed on the
lower line of the ket, while those associated with the corresponding apex
site appear on the upper line. The semicolon separates the consecutive sites
$2i-1$ and $2i$, and $\bigotimes$ denotes the direct product of these local
configurations along the chain. An unspecified spin
label is a free Ising variable, whereas a magnetic label is omitted
for $S=\pm1$ because the projector $1-S^{2}$ switches off that spin
degree of freedom. The two lines in the brace are globally inverted
representatives of the same phase. The states in Eqs.~\eqref{eq:FR2}-\eqref{eq:AF}
follow the same notation.}

The frustrated phase ${\rm FR}_{1}$ has the energy $E_{{\rm FR_{1}}}=-J$
per unit cell and the residual entropy \rev{$\mathcal{S}/(Nk_{\mathrm{B}})=\ln(2)$} reflecting
two-fold degeneracy per unit cell. In the competing regime with higher
values of the electrostatic potential $V/J$ one contrarily finds
the charge-ordered frustrated phase ${\rm FR}_{2}$ displayed in Fig.
\ref{fig:gs}(b) and given by 
\begin{equation}
|{\rm FR}_{2}\rangle=\begin{cases}
\begin{alignedat}{1}\bigotimes^{N/2}_{i=1}|^{\tilde{S}_{2i-1}=0,\tilde{\sigma}_{2i-1}}_{S_{2i-1}=1,}{}^{;\tilde{S}_{2i}=0,\tilde{\sigma}_{2i}}_{;S_{2i}=-1,}\rangle,\\
\bigotimes^{N/2}_{i=1}|^{\tilde{S}_{2i-1}=0,\tilde{\sigma}_{2i-1}}_{S_{2i-1}=-1,}{}^{;\tilde{S}_{2i}=0,\tilde{\sigma}_{2i}}_{;S_{2i}=1,}\rangle,
\end{alignedat}
\end{cases}\label{eq:FR2}
\end{equation}
where the frustrated character of the apex spins follows from a regular
alternation of two charged configurations of the base pseudospins
$+1$ and $-1$. The frustrated phase ${\rm FR}_{2}$ has the energy
$E_{{\rm FR_{2}}}=\Delta-V$ per unit cell and the same residual entropy
\rev{$\mathcal{S}/(Nk_{\mathrm{B}})=\ln(2)$}.

The third frustrated ${\rm FR}_{3}$ phase
with the energy per unit cell $E_{{\rm FR_{3}}}=J-2J_{1}=-J$ is stabilized
under the very special constraint when both magnetic exchange interactions
become equal $J_{1}/J=1$. Within the ${\rm FR}_{3}$ phase, all pseudospins
are frozen at zero $S_{i}=S_{i+1}=\tilde{S}_{i}=0$ and the frustration
arises from the spin configurations $\sigma_{i},\sigma_{i+1},\tilde{\sigma}_{i}$
balancing the competing magnetic interactions $J$ and $J_{1}$ as
shown in Fig. \ref{fig:gs}(c) and given by 
\begin{equation}
|{\rm FR}_{3}\rangle=\bigotimes^{N/2}_{i=1}\bigl|{}^{\tilde{S}_{2i-1}=0,\tilde{\sigma}_{2i-1}}_{S_{2i-1}=0,\sigma_{2i-1}}{}^{;\tilde{S}_{2i}=0,\tilde{\sigma}_{2i}}_{;S_{2i}=0,\sigma_{2i}}\bigr\rangle.
\end{equation}
Each triangular cell admits within the ${\rm FR}_{3}$ phase three
distinct spin configurations $(\sigma_{2i-1},\sigma_{2i},\tilde{\sigma}_{2i-1})=\{(+1,+1,-1),\,(+1,-1,+1),\,(+1,-1,-1)\}$
and its three counterparts obtained upon the local inversion $(\sigma_{2i-1},\sigma_{2i},\tilde{\sigma}_{2i-1})\to(-\sigma_{2i-1},-\sigma_{2i},-\tilde{\sigma}_{2i-1})$.
\rev{Because neighboring triangles share a base spin, these local 
configurations are subject to compatibility constraints. 
Nevertheless, the number of compatible ground states grows 
asymptotically as $W\sim 3^N$, yielding the residual entropy
$\mathcal{S}/(Nk_{\mathrm{B}})=\ln(3)$} due to a threefold local degeneracy of
each triangular cell. The distinction of the ${\rm FR}_{3}$ phase
from the ${\rm FR}_{1}$ phase lies in the enlarged local manifold
of the spin states. Finally, the system also exhibits the antiferromagnetic
(AF) phase, in which the apex spins are antiparallel oriented with
respect to the base spins as depicted in Fig. \ref{fig:gs}(d) and
given by 
\begin{equation}
|{\rm AF}\rangle=\begin{cases}
\begin{alignedat}{1}\bigotimes^{N}_{i=1}|^{\tilde{S}_{i}=0,\tilde{\sigma}_{i}=-1}_{S_{i}=0,\sigma_{i}=1}\rangle,\\
\bigotimes^{N}_{i=1}|^{\tilde{S}_{i}=0,\tilde{\sigma}_{i}=1}_{S_{i}=0,\sigma_{i}=-1}\rangle.
\end{alignedat}
\end{cases}\label{eq:AF}
\end{equation}
The AF phase eliminates the macroscopic degeneracy and hence, it has
null residual entropy and energy \rev{$E_{{\rm AF}}=J-2J_{1}$} per unit
cell.

Fig. \ref{fig:Phase-diagram} displays the ground-state phase diagrams
of the spin-pseudospin sawtooth chain in the $V/J-\Delta/J$ plane
for three representative coupling ratios $J_{1}/J=0.5,\;1.0,\;2.0$.
Four distinct phases can be identified: three frustrated states denoted
as $\mathrm{FR_{1}}$, $\mathrm{FR_{2}}$, and $\mathrm{FR_{3}}$,
and the antiferromagnetic state $\mathrm{AF}$. The background color
encodes the entropy density \rev{$\mathcal{S}/(Nk_{\mathrm{B}})$} at a low
temperature $k_{\mathrm{B}}T/J=0.1$, which highlights the degeneracy
associated with each phase. The regions $\mathrm{FR_{1}}$ and $\mathrm{FR_{2}}$
are characterized by the residual entropy \rev{$\mathcal{S}/(Nk_{\mathrm{B}})=\ln(2)$},
while $\mathrm{FR_{3}}$ shows the higher residual entropy \rev{$\mathcal{S}/(Nk_{\mathrm{B}})=\ln(3)$}.
In contrast, the antiferromagnetic phase $\mathrm{AF}$ is non-degenerate
and thus displays zero residual entropy \rev{$\mathcal{S}/(Nk_{\mathrm{B}})=0$}.

\rev{The colored background in Fig.~\ref{fig:Phase-diagram} is evaluated
at the finite temperature $k_{\mathrm B}T/J=0.1$ and should therefore
not be identified everywhere with the residual entropy. In particular,
near $\Delta/J=0$ several charge configurations have small excitation
gaps and are thermally populated. At exactly $\Delta=\Delta_1=0$ in
the charge-dominated region, $V>\max(J,2J_1-J)$, the zero-temperature
transfer matrix gives the exact residual entropy
$\mathcal S/(Nk_{\mathrm B})=\ln[4(1+\sqrt{2})]\approx 2.2677$. This enhanced
degeneracy accounts for the high entropy close to $\Delta=0$; for
$\Delta>0$ the additional charged configurations are lifted and the
entropy approaches the residual value of the corresponding phase.}

The zero-temperature phase boundaries are determined by level 
crossings between the corresponding ground-state energies. 
At finite temperature, these boundaries evolve into pseudo-critical 
lines described by Eq. (\ref{eq:psd-cond}). For weak coupling $J_{1}/J=0.5$
{[}Fig.\ref{fig:Phase-diagram}(a){]}, the phase diagram is dominated
by extended frustrated domains indicating a high degree of degeneracy
in the spin-pseudospin configurations. As $J_{1}/J$ increases to
unity {[}Fig.\ref{fig:Phase-diagram}(b){]}, the $\mathrm{FR}_{3}$
phase emerges characterized by an additional degeneracy associated
with magnetic polarization along the base sites. In the strong-coupling
limit $J_{1}/J=2.0$ {[}Fig.\ref{fig:Phase-diagram}(c){]}, the system
stabilizes the antiferromagnetic phase, in which the base and apex
spins become ordered with opposite spin polarization, while the pseudospin
degrees of freedom freeze giving rise to a non-degenerate configuration
reminiscent of the magnetically ordered state observed in the simple
cuprate linear chain.

\subsection{Entropy maps and quasi-phases}

We now turn to the analysis of several physical quantities that characterize
the thermodynamic and correlation properties of the spin-pseudospin
sawtooth chain, which serves as a minimal model for 1D cuprates composed
from corner-sharing triangles. The Hamiltonian of this system exhibits
global $\mathbb{Z}_{2}$-symmetries with respect to independent inversions
of both spin and pseudospin variables $\sigma\rightarrow-\sigma$
and $S\rightarrow-S$. Each term in the Hamiltonian either contains
an even number of spins $\sigma$ or is multiplied by the projector
$(1-S^{2})$, while all pseudospin contributions enter only through
quadratic forms such as $S^{2}$, $S_{i}S_{i+1}$ or the mixed product
$S_{3}(S_{1}+S_{2})$. Under the global inversion $S\rightarrow-S$,
these mixed terms remain invariant confirming that the Hamiltonian
preserves the $\mathbb{Z}_{2}$ symmetry.

\rev{In the translation-invariant zero-field Gibbs state used in the
transfer-matrix calculation, these symmetries imply
$\langle\sigma_i\rangle=\langle S_i\rangle
=\langle\tilde{\sigma}_i\rangle
=\langle\tilde{S}_i\rangle=0$. However, symmetry alone does not exclude
phase separation under a global constraint, as in the ferromagnetic
Ising chain at fixed zero magnetization. Such behavior was also found
for a dilute spin-pseudospin chain with excess charge
\citep{yasinskaya2024}. In the present grand-canonical,
charge-symmetric model, no excess charge is imposed, so phase-separated
states do not form additional bulk ground-state regions, although they
may occur at coexistence boundaries.}
Consequently, the relevant physical quantities to be examined below
are restricted to two-point correlation functions and thermodynamic
response functions, whose anomalies provide clear signatures of
pseudo-transition phenomena closely analogous to the charge-spin
interplay observed in $[\mathrm{CuO}]_{\infty}$ cuprate chains.

\begin{figure*}
\includegraphics{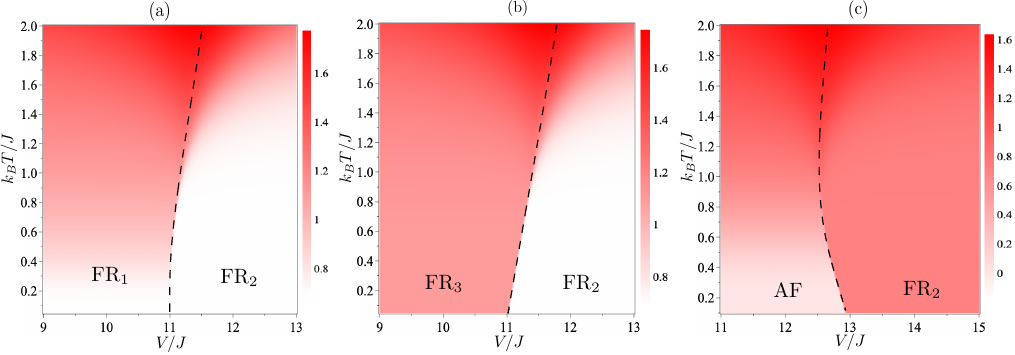}\caption{Density plot of entropy \rev{$\mathcal{S}/(Nk_{\mathrm{B}})$} in the plane $V/J-k_{{\rm B}}T/J$
for $V_{1}=V$, \rev{$\Delta_{1}=\Delta$ and $\Delta/J=10$}, and three representative values of the coupling ratio:
(a) $J_{1}/J=0.5$; (b) $J_{1}/J=1.0$; (c) $J_{1}/J=2.0$. The color
maps illustrate the combined effect of temperature and the coupling
ratio $V/J$ on the \rev{finite-temperature entropy, while the dashed curves show
the pseudo-critical temperature $T_p(V)$ obtained from Eq. (\ref{eq:psd-cond}) and} separate
the quasi-phases.}
\label{fig:S-VT} 
\end{figure*}

Fig. \ref{fig:S-VT} displays the density plot of the entropy \rev{$\mathcal{S}/(Nk_{\mathrm{B}})$}
in the $V/J-k_{\mathrm{B}}T/J$ plane illustrating how temperature
and the coupling ratio $V/J$ influence the entropy landscape and
the appearance of quasi-phases in the spin-pseudospin sawtooth chain.
\rev{Thus, the dashed curves in the three panels also provide a direct
plot of the predicted pseudo-critical temperature as a function of
the electrostatic interaction $V/J$ for the three selected values
of $J_{1}/J$.}
Fig. \ref{fig:S-VT}(a) corresponds to the weak-coupling regime $J_{1}/J=0.5$,
where wide entropy plateaus with \rev{$\mathcal{S}/(Nk_{\mathrm{B}})\approx\ln(2)$}
dominate reflecting the frustrated phases $\mathrm{FR}_{1}$ and $\mathrm{FR}_{2}$.
A narrow nearly vertical ridge following the condition given by Eq.
(\ref{eq:psd-cond}) identifies the pseudo-critical line separating
these quasi-phases. As the temperature increases, this ridge becomes
progressively smoother indicating enhanced thermal mixing and the
fading of pseudo-transition features. The pseudo-transition curve
in this regime can be estimated from Eq. (\ref{eq:prt-Tp-cond}),
which in the low-temperature limit simplifies to $V=\Delta+J+\frac{k_{{\rm B}}T}{2}{\rm e}^{-J/(k_{{\rm B}}T)}+\cdots$
showing that the pseudo-critical line approaches the asymptotic value
$V\to\Delta+J$ as $T\to0$ with an exponentially small positive slope
that captures the onset of the weak spin-charge coupling.

Fig. \ref{fig:S-VT}(b) presents the intermediate coupling case $J_{1}/J=1.0$.
In this regime, the ridge separating the quasi-phases \rev{$\mathrm{FR}_{3}$}
and \rev{$\mathrm{FR}_{2}$} exhibits a positive slope and follows the condition
given by Eq.~(\ref{eq:psd-cond}). These phases are characterized
\rev{at low temperature by entropy plateaus approaching the exact
residual values $\mathcal{S}/(Nk_{\mathrm{B}})=\ln3$ and
$\mathcal{S}/(Nk_{\mathrm{B}})=\ln2$, respectively.} This
regime reflects a balance between spin and charge correlations, analogous
to the coexistence of $\mathrm{Cu}^{2+}$ and mixed-valence $\mathrm{Cu}^{1+}$/$\mathrm{Cu}^{3+}$
configurations in $[\mathrm{CuO}]_{\infty}$ chains. The pseudo-transition
curve can be obtained from Eq.~(\ref{eq:prt-Tp-cond}), which in
the low-temperature limit reduces to $V=\Delta+J+k_{{\rm B}}T\ln(\frac{3}{2})-\frac{5}{3}k_{{\rm B}}T{\rm e}^{-\Delta/k_{{\rm B}}T}+\cdots$
indicating that the pseudo-critical line departs linearly from the
asymptotic value $V=\Delta+J$ with a weak positive slope proportional
to $\ln(3/2)$, which serves as a clear signature of the intermediate
spin--charge coupling regime.

Fig. \ref{fig:S-VT}(c) depicts the strong-coupling limit $J_{1}/J=2.0$,
where the entropy plateaus are sharply bounded by a line with negative
slope obeying the same condition given by Eq. (\ref{eq:psd-cond}).
This ridge separates the antiferromagnetic phase $\mathrm{AF}$ characterized
by \rev{a low-temperature entropy plateau approaching the exact residual
value $\mathcal{S}/(Nk_{\mathrm{B}})=0$} from
the frustrated phase $\mathrm{FR}_{2}$ with \rev{the exact residual entropy
$\mathcal{S}/(Nk_{\mathrm{B}})=\ln(2)$}. The suppression of entropy
in the $\mathrm{AF}$ region indicates that magnetic exchange interactions
dominate over charge fluctuations leading to a configuration reminiscent
of magnetically ordered segments in quasi-one-dimensional cuprates.
In this strong-coupling regime, the pseudo-transition curve follows
from Eq. (\ref{eq:prt-Tp-cond}), which in the low-temperature regime
simplifies to $V=\Delta+3J-k_{{\rm B}}T\ln(2)+2k_{{\rm B}}T{\rm e}^{-2J/k_{{\rm B}}T}+\cdots$
showing that the pseudo-critical line approaches the asymptotic value
$V=\Delta+3J$ as $T\to0$. The negative slope proportional to $-\ln(2)$
reflects the dominance of magnetic exchange interactions and the progressive
suppression of charge degrees of freedom in the antiferromagnetic
regime.

\section{Thermodynamic and correlation properties}

\label{sec:Results}

Having established the pseudo-critical condition, we now analyze the
thermodynamic behavior of the spin-pseudospin sawtooth chain. Particular
attention is paid to typical temperature dependence of the entropy
and specific heat, whose pronounced anomalies signal the pseudo-transition
lines predicted in Sec. \ref{subsec:psd-cirt-cond}. Fig. \ref{fig:S-VT}(a-c)
illustrate how these anomalies manifest in different coupling regimes
reproducing the characteristic behavior of pseudo-transitions already
reported in several one-dimensional models \citep{galisova2015,rojaspseudo,universal,rojase,hutak2021,yin2024,yin2024prb}

\subsection{Entropy and specific heat}

\begin{figure*}
\includegraphics[scale=0.95]{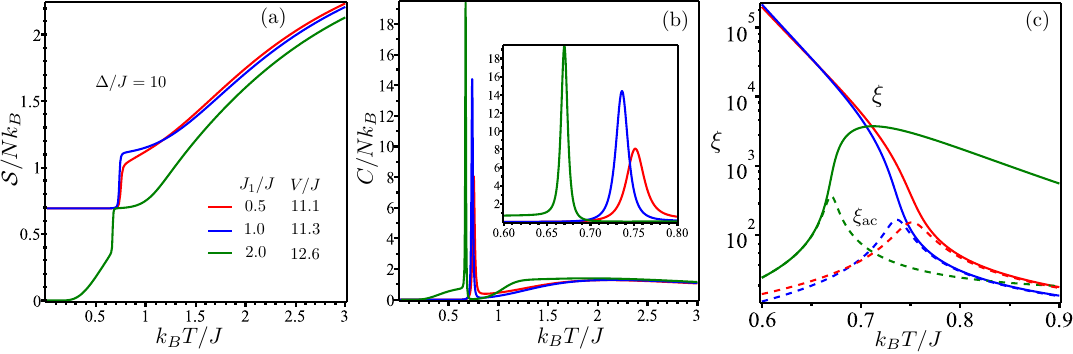}\caption{(a) Entropy as a function of temperature for a fixed value of $\Delta/J=10.0$
\rev{with $\Delta_1=\Delta$ and $V_1=V$}, and three representative sets of interaction parameters: $V/J=11.1$
and $J_{1}/J=0.5$ (red curve); $V/J=11.3$ and $J_{1}/J=1.0$ (blue
curve); $V/J=12.6$ and $J_{1}/J=2.0$ (green curve). (b) Temperature
variations of the specific heat for the same set of parameters {[}color
code is identical as in the panel (a){]}; (c) Solid curves show the correlation length $\xi$
as a function of temperature for the same set of parameters as in
the panel (a), \rev{while the dashed curves show the avoided-crossing
quantity $\xi_{\rm ac}$.}}
\label{fig:Entr-C-xi} 
\end{figure*}

Fig. \ref{fig:Entr-C-xi} presents the temperature dependence of three
key thermodynamic quantities - the entropy, the specific heat, and
the correlation length, which provide direct evidence of a pseudo-transition
in the spin-pseudospin sawtooth chain. This behavior closely resembles
that observed in the minimal spin-pseudospin model for $[\mathrm{CuO}]_{\infty}$
cuprate chains \citep{strecka_epjb}, where a competition between
magnetic and charge degrees of freedom gives rise to a sharp yet continuous
pseudo-critical crossover between two distinct short-range ordered
regimes.

Fig. \ref{fig:Entr-C-xi}(a) shows the entropy density \rev{$\mathcal{S}/(Nk_{\mathrm{B}})$}
as a function of temperature for three representative parameter sets
each with a fixed value of the parameter $\Delta/J=10.0$. The red
curve corresponds to \rev{$V/J=V_1/J=11.1$}, and $J_{1}/J=0.5$. The entropy
exhibits a sudden but continuous rise at $k_{\mathrm{B}}T_{p_{1}}/J=0.75345587$
coinciding with the pseudo-critical condition given by Eq. (\ref{eq:psd-cond}).
The blue curve corresponding to \rev{$V/J=V_1/J=11.3$} and $J_{1}/J=1.0$
shows a similar sharp increase at $k_{\mathrm{B}}T_{p_{2}}/J=0.7372288$,
while the green curve for \rev{$V/J=V_1/J=12.6$} and $J_{1}/J=2.0$ exhibits
an analogous pseudo-critical behavior at $k_{\mathrm{B}}T_{p_{3}}/J=0.6703250$.
In all cases, the abrupt entropy change reflects a rapid release of
the residual entropy signaling the pseudo-transition between the corresponding
\rev{pairs of quasi-phases: $\mathrm{FR}_{2}$-$\mathrm{FR}_{1}$,
$\mathrm{FR}_{2}$-$\mathrm{FR}_{3}$, and $\mathrm{FR}_{2}$-AF,
respectively. The quoted values of $T_{p_1}$, $T_{p_2}$, and $T_{p_3}$
are obtained by solving the exact implicit condition
Eq.~(\ref{eq:psd-cond}) for the three parameter sets. Equation
(\ref{eq:prt-Tp-cond}) gives their low-temperature analytical estimates.
Because Fig.~\ref{fig:corlns} uses the same parameter sets, these
temperatures also locate its correlation-function anomalies.}

The residual entropies of the ground-state phases, discussed in Sec.
\ref{subsec:sec.3.A}, are given by $\mathcal{S}^{{\rm FR}_{1}}/(Nk_{\mathrm{B}})=\ln2$,
$\mathcal{S}^{{\rm FR}_{2}}/(Nk_{\mathrm{B}})=\ln2$, and $\mathcal{S}^{{\rm FR}_{3}}/(Nk_{\mathrm{B}})=\ln3$.
As the temperature approaches the pseudo-critical point, configurations
associated with different sectors contribute simultaneously to the
partition function, leading to a rapid but continuous variation of
the entropy, as observed in Fig. \ref{fig:Entr-C-xi}(a). In particular,
the absence of an increasing residual entropy at the phase boundary
is a characteristic signature of pseudo-transitions, which arise from
the competition between sectors with different degeneracies.

Following the criterion that pseudo-transitions are governed by competing
sectors whose residual entropies constrain the thermodynamic response
\citep{spc-crit}, one finds for the crossover between the phases
$\mathrm{FR}_{2}$ and $\mathrm{FR}_{3}$ that
\begin{equation}
\ln2\le\mathcal{S}_{0}^{c}/(Nk_{\mathrm{B}})\le\ln3,\qquad\rev{\mathcal{S}_{0}^{c}/(Nk_{\mathrm{B}})=\ln\sqrt{6}},
\end{equation}
while for the crossover between $\mathrm{FR}_{2}$ and AF,
\begin{equation}
0\le\mathcal{S}_{0}^{c}/(Nk_{\mathrm{B}})\le\ln2,\qquad\rev{\mathcal{S}_{0}^{c}/(Nk_{\mathrm{B}})=\ln\sqrt{2}}.
\end{equation}
For the crossover between $\mathrm{FR}_{1}$ and $\mathrm{FR}_{2}$,
the condition is trivially satisfied since 
\begin{equation}
\mathcal{S}^{{\rm FR}_{1}}/(Nk_{\mathrm{B}})=\mathcal{S}^{{\rm FR}_{2}}/(Nk_{\mathrm{B}})=\ln2.
\end{equation}
These results are consistent with the entropy behavior displayed in
Fig. \ref{fig:Entr-C-xi}, and further support the interpretation
of the pseudo-transition as a consequence of the competition between
sectors with distinct degeneracies. 

\rev{According to Ref.~\citep{spc-crit}, $\mathcal S_{0}^{c}$ is the
interface residual entropy obtained in the limit $T\to0$ taken along
the pseudo-critical line. For two competing sectors of local degeneracies
$g_a$ and $g_b$, it is
$\mathcal S_{0}^{c}/(Nk_{\mathrm B})=\ln\sqrt{g_ag_b}$ and therefore
lies between their residual entropies. This inequality is a necessary,
but not sufficient, diagnostic of a pseudo-transition; the quasi-phase
boundary is determined independently by Eq.~(\ref{eq:psd-cond}).}

Fig. \ref{fig:Entr-C-xi}(b) displays thermal variations of the specific
heat $C/Nk_{\mathrm{B}}$ for the same parameter sets. The extremely
sharp yet finite peaks observed at $T_{p_{1}}$, $T_{p_{2}}$, and
$T_{p_{3}}$ are characteristic signatures of pseudo-transitions.
These peaks closely resemble the narrow specific-heat anomalies reported
in $[\mathrm{CuO}]_{\infty}$ chains near the crossover between charge-ordered
and antiferromagnetic phases \citep{strecka_epjb}. The height and
sharpness of these peaks are strongly influenced by the deviation
of the interaction parameters $J_{1}/J$ and $V/J$ from the respective
phase boundaries depicted in the ground-state phase diagram shown
in Fig. \ref{fig:Phase-diagram}, whereas the robustness of the pseudo-transition
is also highly sensitive with respect to increasing temperature.

Fig. \ref{fig:Entr-C-xi}(c) depicts the temperature dependence of
the correlation length \rev{defined earlier as $\xi=1/\ln(\frac{\lambda_{1}}{\lambda_{\rm sub}})$
under the same conditions, while the avoided-crossing quantity is given
by $\xi_{\rm ac}=1/\ln(\frac{\lambda_{1}}{\lambda_{2}})$}. In full correspondence with the specific
heat, $\xi_{\rm ac}$ displays a gigantic and sharply localized maximum precisely
at the pseudo-critical temperature identified in panels (a) and (b). 
\rev{The corresponding pseudo-critical temperatures are
$k_{\mathrm B}T_{p_1}/J=0.75345587$,
$k_{\mathrm B}T_{p_2}/J=0.7372288$, and
$k_{\mathrm B}T_{p_3}/J=0.6703250$.}
This behavior confirms that the pseudo-transition is accompanied by
a sudden yet finite enhancement of short-range correlations similar
to that observed in one-dimensional $[\mathrm{CuO}]_{\infty}$ cuprates
\citep{strecka_epjb}. Altogether, the results presented in Fig. \ref{fig:Entr-C-xi}
establish a consistent thermodynamic picture: entropy, specific heat,
and avoided crossing curves exhibit abrupt but continuous sharp anomalies
clearly demonstrating the pseudo-critical behavior.

\subsection{Local correlations}

The analysis of local and pair correlations offers complementary insight
into the cooperative behavior of spins and pseudospins within the
sawtooth chain, which serves as a minimal representation of cuprate
chain consisting of corner-sharing triangles. In this framework, the
spin variables $\sigma_{i}$ and pseudospin variables $S_{i}$ are
connected with the base sites of the sawtooth chain, while the spin
variables $\tilde{\sigma}_{i}$ and the pseudospin variables $\tilde{S}_{i}$
are assigned to the apex sites at the top of each triangular unit
cell. The geometric asymmetry of the sawtooth chain naturally introduces
a hierarchy of coupling constants $(J,J_{1})$ for the magnetic exchange
interactions and $(V,V_{1})$ for the electrostatic interactions,
which mediate both base-base and apex-base correlations. As in the
analogous model of one-dimensional cuprate chains $[\mathrm{CuO}]_{\infty}$
\citep{strecka_epjb}, these correlations can be obtained directly
from derivatives of the free energy per unit cell $f=-\frac{1}{\beta}\ln(\lambda_{1})$
with respect to the coupling parameters appearing in the Hamiltonian
(\ref{eq:H-cell}).

The quadratic terms $\tfrac{\Delta}{2}(S^{2}_{i}+S^{2}_{i+1})$ and
$\Delta_{1}\tilde{S}^{2}_{i}$ in the Hamiltonian (\ref{eq:H-cell})
correspond to the on-site or autocorrelation functions that quantify
the probability of charge occupancy on the base and apex sites, respectively.
Hence, the respective thermal averages can be obtained as $\langle S^{2}_{i}\rangle=\frac{\partial f}{\partial\Delta}$
and $\langle\tilde{S}^{2}_{i}\rangle=\frac{\partial f}{\partial\Delta_{1}}$.
These quantities characterize how charge fluctuations are distributed
among the base and apex sites.

The nearest-neighbor correlation between two base pseudospins is obtained
by differentiating the free energy with respect to the electrostatic
potential $V$: $\langle S_{i}S_{i+1}\rangle=\frac{\partial f}{\partial V}$.
This quantity determines the electrostatic correlations between adjacent
base sites of the sawtooth chain. The electrostatic coupling between
an apex pseudospin $\tilde{S}_{i}$ and its two adjacent base pseudospins
$S_{i}$ and $S_{i+1}$ is governed by the electrostatic potential
$V_{1}$. The corresponding correlation functions follow from differentiation
with respect to $V_{1}$: $\langle\tilde{S}_{i}S_{i}\rangle=\langle\tilde{S}_{i}S_{i+1}\rangle=\frac{\partial f}{2\partial V_{1}}$.
These correlations quantify how charged configurations on the base
and apex sites are mutually correlated.

The magnetic sector exhibits an analogous behavior. Differentiation
of the free energy with respect to the exchange parameter $J$ gives
the correlation between two neighboring base spins $\langle(1-S^{2}_{i})(1-S^{2}_{i+1})\,\sigma_{i}\sigma_{i+1}\rangle=\frac{\partial f}{\partial J}$.
In this model, the projectors $(1-S^{2}_{i})(1-S^{2}_{i+1})$ restrict
the correlation to the sector $S_{i}=S_{i+1}=0$. 
Since the inactive spin labels associated with $S_i=\pm1$ occur 
symmetrically in the trace and cancel, the projected correlation 
is equal to the simplified average $\langle\sigma_{i}\sigma_{i+1}\rangle=\frac{\partial f}{\partial J}$.
Similarly, differentiation with respect to $J_{1}$ yields the apex-base
magnetic correlation $\langle(1-\tilde{S}^{2}_{i})(1-S^{2}_{i})\sigma_{i}\tilde{\sigma}_{i}\rangle=\frac{\partial f}{2\partial J_{1}}$,
which reduces to $\langle\sigma_{i}\tilde{\sigma}_{i}\rangle=\frac{\partial f}{2\partial J_{1}}$.

\begin{figure*}
\includegraphics[scale=0.95]{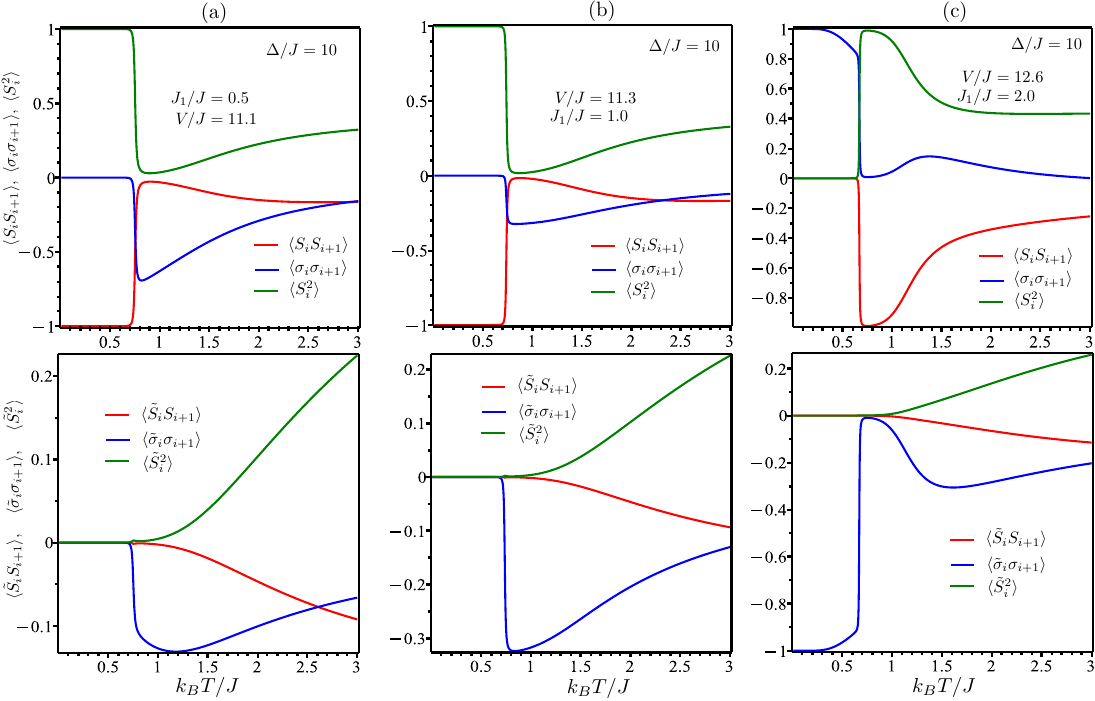}\caption{Temperature dependence of spin and pseudospin correlation functions
for a fixed value of $\Delta/J=10.0$ \rev{with $\Delta_1=\Delta$ and $V_1=V$} and three selected sets of the
interaction parameters: (a), (d) $V/J=11.1$ and $J_{1}/J=0.5$; (b),
(e) $V/J=11.3$ and $J_{1}/J=1.0$; (c), (f) $V/J=12.6$ and $J_{1}/J=2.0$.
Top panels display correlations on the base sites $\langle S_{i}S_{i+1}\rangle$
(red), $\langle\sigma_{i}\sigma_{i+1}\rangle$ (blue), and $\langle S^{2}_{i}\rangle$
(green), while bottom panels show apex-base correlations $\langle \tilde{S}_{i}S_{i+1}\rangle$
(red), $\langle\tilde{\sigma}_{i}\sigma_{i+1}\rangle$ (blue), and
$\langle\tilde{S}^{2}_{i}\rangle$ (green).}
\label{fig:corlns} 
\end{figure*}

In Fig.~\ref{fig:corlns} we report the temperature dependence of
several nearest-neighbor and on-site correlation functions, which
together illustrate the microscopic origin of the pseudo-transition
in the spin-pseudospin sawtooth chain. These correlations capture
the interplay between charge and magnetic degrees of freedom distributed
over the triangular plaquettes of the sawtooth lattice, an arrangement
reminiscent of the corner-sharing geometry of some cuprate chains
\citep{delafossite1,delafossite2,delafossite3,delafossite4,delafossite5}.

Fig.~\ref{fig:corlns}(a) shows the results for $\Delta/J=10.0$,
\rev{$V/J=V_{1}/J=11.1$}, and $J_{1}/J=0.5$. The top panel compares the nearest-neighbor
correlation between base pseudospins $\langle S_{i}S_{i+1}\rangle$
(red curve), the magnetic correlation between base spins $\langle\sigma_{i}\sigma_{i+1}\rangle$
(blue curve), and the on-site pseudospin autocorrelation $\langle S^{2}_{i}\rangle$
(green curve). Below the pseudo-critical temperature $k_{\mathrm{B}}T_{p_{1}}/J=0.75345587$,
all correlations remain nearly constant indicating a frozen frustrated
configuration within the quasi-phase $\mathrm{FR}_{2}$. As the temperature
crosses $T_{p_{1}}$, a pronounced but continuous change emerges:
$\langle S^{2}_{i}\rangle$ increases sharply revealing the activation
of local charge states, while $\langle S_{i}S_{i+1}\rangle$ and $\langle\sigma_{i}\sigma_{i+1}\rangle$
show more moderate variations signaling the onset of new local correlations.
The bottom panel shown in Fig.~\ref{fig:corlns}(d) displays the
correlations involving the apex sites $\langle \tilde{S}_{i}S_{i+1}\rangle$
(red), $\langle\tilde{\sigma}_{i}\sigma_{i+1}\rangle$ (blue), and
$\langle\tilde{S}^{2}_{i}\rangle$ (green). Below $T_{p_{1}}$, all
these quantities remain nearly zero reflecting the decoupled nature
of the apex units. At $T_{p_{1}}$, the autocorrelation $\langle\tilde{S}^{2}_{i}\rangle$
exhibits a pronounced increase, while the nearest-neighbor correlations
$\langle \tilde{S}_{i}S_{i+1}\rangle$ and $\langle\tilde{\sigma}_{i}\sigma_{i+1}\rangle$
display only small but finite increase indicating a weak hybridization
between base and apex degrees of freedom induced by the pseudo-transition.

Fig.~\ref{fig:corlns}(b) and (e) correspond to the parameter set
$\Delta/J=10.0$, \rev{$V/J=V_{1}/J=11.3$}, and $J_{1}/J=1.0$. The overall
behavior is qualitatively similar to that previously discussed in
panel (a), but the anomalies become sharper and more pronounced around
the pseudo-critical temperature $k_{\mathrm{B}}T_{p_{2}}/J=0.7372288$.
The base correlations, particularly $\langle\sigma_{i}\sigma_{i+1}\rangle$,
exhibit enhanced sensitivity to temperature and display an abrupt
change at $T_{p_{2}}$, which marks the crossover between the quasi-phases
$\mathrm{qFR}_{2}$ and $\mathrm{qFR}_{3}$. In contrast, the correlations
involving apex sites only exhibit small variations indicating that
the pseudo-transition remains predominantly controlled by interactions
within the base subsystem.

Fig.~\ref{fig:corlns}(c) and (f) show the correlations for the strong-coupling
regime $\Delta/J=10.0$, \rev{$V/J=V_{1}/J=12.6$}, and $J_{1}/J=2.0$. The
overall trends are consistent with those in panels (a) and (b): below
the pseudo-critical temperature $T_{p_{3}}/J=0.6703250$ all correlations
remain nearly constant, while distinct abrupt anomalies appear at
$T_{p_{3}}$ in both magnetic and pseudospin correlations. As the
interaction ratio $J_{1}/J$ increases, these anomalies become sharper
and more localized revealing enhanced coherence of frustrated configurations
and a stronger coupling between the base and apex subsystems.

Altogether, the results presented in Fig. \ref{fig:corlns} demonstrate
that the pseudo-transition in the spin-pseudospin sawtooth chain manifests
microscopically as a sudden yet continuous rearrangement of local
correlations along the triangular plaquettes. The pronounced variations
of $\langle S^{2}_{i}\rangle$, $\langle S_{i}S_{i+1}\rangle$, and
$\langle\sigma_{i}\sigma_{i+1}\rangle$ signal a crossover from a
charge-dominated frustrated regime to a more magnetically correlated
configuration reflecting the cooperative spin-charge interplay characteristic
of cuprate chains of corner-sharing triangles \citep{delafossite1,delafossite2,delafossite3,delafossite4,delafossite5}.

\section{Conclusion}

\label{sec:Conclusion}

In the present work, we have investigated the thermodynamic properties
of a spin-pseudospin sawtooth chain, which may be regarded as a minimal
representation of cuprate chains of corner-sharing triangles. The
model incorporates both magnetic exchange and electrostatic interactions
between base and apex units leading to a rich interplay between spin
and charge degrees of freedom. By means of the transfer-matrix approach,
we have derived closed-form analytical expressions for the Boltzmann
weights, the eigenvalue spectrum, and the pseudo-critical condition
determining the location of thermodynamic anomalies. In addition,
we obtained exact results for the entropy, specific heat, correlation
length, avoided crossing, and distinct local correlation functions.

The ground-state analysis revealed the antiferromagnetic phase AF
and three frustrated phases $\mathrm{FR}_{1}$, $\mathrm{FR}_{2}$,
$\mathrm{FR}_{3}$ distinguished by characteristic residual entropies.
Exact results demonstrate that the zero-temperature phase boundaries
lead at finite temperatures into narrow entropy ridges, which trace
the pseudo-critical line given by the analytical condition $w_{1}+w_{5}=(w_{3}+w_{4})/2$.
The entropy, specific heat, \rev{and avoided-crossing scale display abrupt
 yet continuous changes along this line, signaling clear signatures
of pseudo-transitions, while the physical 
 correlation length reflects the dominant spatial correlation channel}.

A detailed analysis of the correlation functions confirms that the
pseudo-transition corresponds microscopically to a collective rearrangement
of local spin and pseudospin correlations. Below the pseudo-critical
temperature, the system resides in a charge-dominated frustrated regime,
while magnetic correlations become dominant above it. The pronounced
sensitivity of both thermodynamic and microscopic observables to the
coupling ratio $J_{1}/J$ highlights the role of geometric frustration
inherent to the sawtooth chain geometry. In particular, approaching
the zero-temperature phase boundary enhances the coherence of correlations
and leads to sharper pseudo-transition features analogous to those
observed in several one-dimensional systems.

In summary, the spin-pseudospin sawtooth chain captures the essential
mechanism by which spin-charge competition and lattice geometry jointly
give rise to pseudo-transition behavior at finite temperatures. This
study thus extends the framework established for one-dimensional \rev{spin-pseudospin}
model to a geometrically frustrated lattice providing further insight
into the thermodynamic signatures of short-range order in cuprate
chains of corner-sharing triangles \citep{delafossite1,delafossite2,delafossite3,delafossite4,delafossite5}.
Future investigations may address the effects of external magnetic
or electric fields, as well as the relaxation of symmetry constraints
to explore possible analogues of charge-spin separation and emergent
coherence in related low-dimensional systems.
\begin{acknowledgments}
J.S. acknowledges the financial support by the grant of Slovak Research
and Development Agency under the contract No. APVV-24-0091 and the
grant of The Ministry of Education, Research, Development and Youth
of the Slovak Republic under the contract No. VEGA 1/0695/23. O.R.
acknowledges partial financial support from the Brazilian agencies
CNPq, CAPES, and FAPEMIG.
\end{acknowledgments}


\begin{thebibliography}{99}
\bibitem{mattis} D.~C.~Mattis, \textit{The Many-Body Problem: An
Encyclopedia of Exactly Solved Models in One Dimension} (World Scientific,
Singapore, 1993).

\bibitem{vanhove} L.~van Hove, Physica \textbf{16}, 137 (1950).

\bibitem{cuesta} J.~A.~Cuesta and A.~Sánchez, J. Stat. Phys. \textbf{115},
869 (2004).

\bibitem{galisova2015} L.~Gálisová and J.~Strečka, Phys. Rev. E
\textbf{91}, 022134 (2015).

\bibitem{rojaspseudo} S.~M.~de Souza and O.~Rojas, Solid State
Commun. \textbf{269}, 131 (2018).

\bibitem{universal} O.~Rojas, J.~Strečka, M.~L.~Lyra, and S.~M.~de
Souza, Phys. Rev. E \textbf{99}, 042117 (2019).

\bibitem{rojase} O.~Rojas, Braz. J. Phys. \textbf{50}, 675 (2020).

\bibitem{hutak2021} T.~Hutak, T.~Krokhmalskii, O.~Rojas, S.~M.~de~Souza,
and O.~Derzhko, Phys. Lett. A \textbf{387}, 127020 (2021).

\bibitem{yin2024} W.~Yin, Phys. Rev. Res. \textbf{6}, 013331 (2024).

\bibitem{yin2024prb} W.~Yin, Phys. Rev. B \textbf{109}, 214413 (2024).

\bibitem{yin-tsvelik-prl-2024} W.~Yin and A.~M.~Tsvelik, Phys.
Rev. Lett. \textbf{133}, 266701 (2024).

\bibitem{yin-prb-2026} W.~Yin, Phys. Rev. B \textbf{113}, 104418
(2026).

\bibitem{rojas-pre-2021} O.~Rojas, S.~M.~de Souza, J.~Torrico,
L.~M.~Veríssimo, M.~S.~S.~Pereira, and M.~L.~Lyra, Phys. Rev.
E \textbf{103}, 042123 (2021).

\bibitem{rojas-pre-2024} O.~Rojas, S.~M.~de Souza, J.~Torrico,
L.~M.~Veríssimo, M.~S.~S.~Pereira, M.~L.~Lyra, and O.~Derzhko,
Phys. Rev. E \textbf{110}, 024130 (2024).

\bibitem{strecka_epjb} J.~Strečka and K.~Karl'ová, Eur. Phys. J.
B \textbf{97}, 74 (2024).

\bibitem{moskvin1} A.~S.~Moskvin, Phys. Rev. B \textbf{84}, 075116
(2011).

\bibitem{moskvin2} A.~S.~Moskvin, J. Phys.: Condens. Matter \textbf{25},
085601 (2013).

\bibitem{panov} Y.~D.~Panov \textit{et al.}, J. Magn. Magn. Mater.
\textbf{477}, 162 (2019).

\bibitem{delafossite1} R. J. Cava, H. W. Zandbergen, A. P. Ramirez,
H. Takagi, C. T. Chen, J. J. Krajewski, W. F. Peck, J. V. Waszczak,
G. Meigs, R. Roth, L. F. Sand Schneemeyer, J. Solid State Chem. \textbf{104},
437 (1993).

\bibitem{delafossite2} G. Van Tendeloo, O. Garlea, C. Darie, C. Bougerol-Chaillout,
and P. Bordet, J. Solid State Chem. \textbf{156}, 428 (2001).

\bibitem{delafossite3} V. Simonet, R. Ballou, A. P. Murani, O. Garlea,
C. Darie, and P. Bordet, J. Phys.: Condens. Matter \textbf{16}, S805
(2004).

\bibitem{delafossite4} S. A. Blundell and M. D. Núnez-Regueiro, J.
Phys.: Condens. Matter \textbf{16}, S791 (2004).

\bibitem{delafossite5} S. Capponi, C. Lacroix, O. Le Bacq, A. Pasturel,
and M. D. Núnez-Regueiro, J. Phys.: Condens. Matter \textbf{19}, 145233
(2007).

\bibitem{baxt82} R.~J.~Baxter, \textit{Exactly solved models in
statistical mechanics} (Academic Press, New York, 1982).

\bibitem{spc-crit}O. Rojas, Physica A \textbf{685}, 131295 (2026) 

\bibitem{yasinskaya2024}\rev{D.~Yasinskaya and Y.~D.~Panov, Phys. Rev. E
\textbf{110}, 044118 (2024).}

\end{thebibliography}
\end{document}